# Economic complexity and inequality at the national and regional level

**Dominik Hartmann[1] and Flávio L. Pinheiro[2]**

1 Federal University of Santa Catarina (UFSC), Brazil

2 Nova Information Management School (NOVA IMS), Universidade Nova de Lisboa, Portugal

Prepared as a chapter for the HANDBOOK OF COMPLEXITY ECONOMICS
being edited by P. Chen, W. Elsner, and A. Pyka

**Abstract.** Recent studies have found evidence of a negative association between economic complexity and inequality at the country level. Moreover, evidence suggests that sophisticated economies tend to outsource products that are less desirable (e.g. in terms of wage and inequality effects), and instead focus on complex products requiring networks of skilled labor and more inclusive institutions. Yet the negative association between economic complexity and inequality on a coarse scale could hide important dynamics at a fine-grained level. Complex economic activities are difficult to develop and tend to concentrate spatially, leading to "winner-take-most" effects that spur regional inequality in countries. Large, complex cities tend to attract both high- and low-skills activities and workers, and are also associated with higher levels of hierarchies, competition, and skill premiums. As a result, the association between complexity and inequality reverses at regional scales; in other words, more complex regions tend to be more unequal. Ideas from polarization theories, institutional changes, and urban scaling literature can help to understand this paradox, while new methods from economic complexity and relatedness can help identify inclusive growth constraints and opportunities.

**Acknowledgments**. We would like to thank César Hidalgo, Mayra Bezerra, Miguel Guevara, Cristian Jara-Figueroa, Pierre-Alex Balland, Mary Kaltenberg, Andreas Pyka, and Diogo Ferraz for the multiple discussions on economic complexity and inequality over the years that contributed to shaping the ideas discussed in this chapter. DH expresses his gratitude for the financial support of CNPq (406943/2021-4 and 315441/2021-6). FLP is thankful for the financial support provided by FCT Portugal under the project UIDB/04152/2020 — Centro de Investigação em Gestão de Informação (MagIC).

# 1 Introduction

Academic literature has yet to agree on the relationship between economic development and inequality. More than half a century ago, Simon Kuznets (1955) argued that during the process of economic development, the income inequality of countries first rises, but then falls again at higher stages of development. However, more recent authors expect inequality to increase with the division of wealth, power, and labor (Hodgson, 2003). Additionally, (neo-)structuralist ideas and world-system theories point to core-periphery (semi-periphery) structures involving countries and regions participating in the global economy (Bielschowsky, 2009: Arenti e Filomeno, 2007). Yet this lack of agreement may be partly due to significant differences in what different authors consider to be valid measures of economic development and inequality. Moreover, it may also be due to whether the focus of the analysis is on the national or regional level.

Due to data availability and (relative) simplicity of the concept, the most widely adopted measures of economic development have focused on capturing the expansion of aggregate production (e.g., GDP or GDP per capita); in other words, economic growth. Recent advances in quantitative methods, though, have also allowed for a depiction of the quality of economic development. This includes analyzing what type of products and services economies can produce, and thus depict the knowledge that is embedded in economies and societies (Hidalgo et al., 2007, Hidalgo and Hausmann, 2009, Hausmann et al., 2014; Hidalgo 2015, Hidalgo 2021). It makes a difference for the inclusive growth prospects of an economy if it focuses on simple products that are based on natural resources, cheap labor, or economies of scale (such as crude petroleum, textile industries, or cocoa beans); or, if instead, it focuses on a variety of complex products based on high knowledge intensity, networks of skilled labor, or collective learning (such as cars, robots, and medicine) (Hartmann et al., 2017; Ferraz et al., 2021. In this regard, measures like the Economic Complexity Index (ECI), Fitness Index, or ECI+ have used information on the diversity and ubiquity of products that economies export to infer their productive capabilities and thus the quality of economic development (Hidalgo and Hausmann, 2009; Tacchella et al., 2012; Hausmann et al., 2014; Gao and Zhou, 2018; Albeaik, 2017).

It must be noted that the impact of economic complexity on inequality may also depend on the spatial level of analysis (such as cities versus countries) as well as the definition of inequality, such as income inequality (Milanovic, 2012; Azzoni, 2001), poverty (World Bank, 2020), or human development (Sen, 1999; UNDP, 1990). Recent economic complexity research at the country level has shown that countries that export more complex products tend to have lower levels of income inequality, more inclusive institutions, and higher levels of human development (Hartmann et al., 2017, Ferraz et al., 2021). For instance, Slovenia, South Korea, and Germany have lower levels of income inequality than Saudi Arabia, South Africa, and Brazil. However, research on the regional level has also pointed to a reverse association in which complex regions tend to have higher levels of income inequality (Sbardella et al., 2017; Marco et al., 2022; Heinrich Mora et al., 2021). For instance, high levels of inequality can be observed in large complex cities, such as New York, San Francisco, or Sao Paulo. Thus, the early evidence of this emergent field of research suggests a Simpson's Paradox, one in which the sign of the relationship between development and inequality reverses on different aggregation levels.

As Figure 1 illustrates, while the Economic Complexity Index (ECI) has a negative association with the GINI index at the country level, it has a slightly positive association at the metropolitan areas level in the US and the mesoregion level in Brazil. We argue here that there are three key reasons for this behavior: (1.) At the country level, inclusive institutions tend to co-evolve positively with higher levels of economic complexity (Engerman and Sokoloff, 1997; Hartmann et al., 2017). The competitive export of sophisticated products tends to require more skilled and well-paid labor and institutions that allow for mutual learning between different agents of the economy. Moreover, (2.) highly complex economies tend to outsource undesirable products, such as simple textile industries, that are mainly based on cheap labor or resource exploitation. However, (3.) at the regional level, spatial agglomeration effects and the co-existence between simple and complex activities in large cities gain prominence. Migration, innovation, and labor market polarization effects can lead to high levels of inequality within and across cities/regions (Hidalgo et al., 2007; Balland et al., 2020; Hartmann et al., 2020, Pinheiro et al. 2022). These dynamics have also been widely studied in urban scaling literature, which analyzed how the dimensions of cities and metropolitan areas affect their outputs and how a country's structural organization is conditioned by the complex interplay between urban areas (Bettencourt et al., 2010; Rybski et al., 2019). This literature has documented the benefits and caveats of agglomeration in terms of human outputs (Bettencourt and West, 2010), its operating costs (Curado et al., 2021), and productive structures (Youn et al., 2016), showing clear evidence of the rising inequality as the side product of urban agglomeration (Heinrich Mora et al., 2021).

This chapter reviews the association between economic development and inequality from an economic complexity perspective (Hidalgo et al., 2007, Hidalgo and Hausmann, 2009; Hausmann et al., 2014, Hartmann, 2014; Hartmann et al., 2017; Hidalgo, 2021). Moreover, it makes use of insights from the literature in development economics, economic geography, and urban scaling to understand the paradox of a reverse effect on the national and regional level. Finally, it discusses how new methods from relatedness and complexity research can help identify constraints and opportunities for inclusive growth and discusses policy considerations.

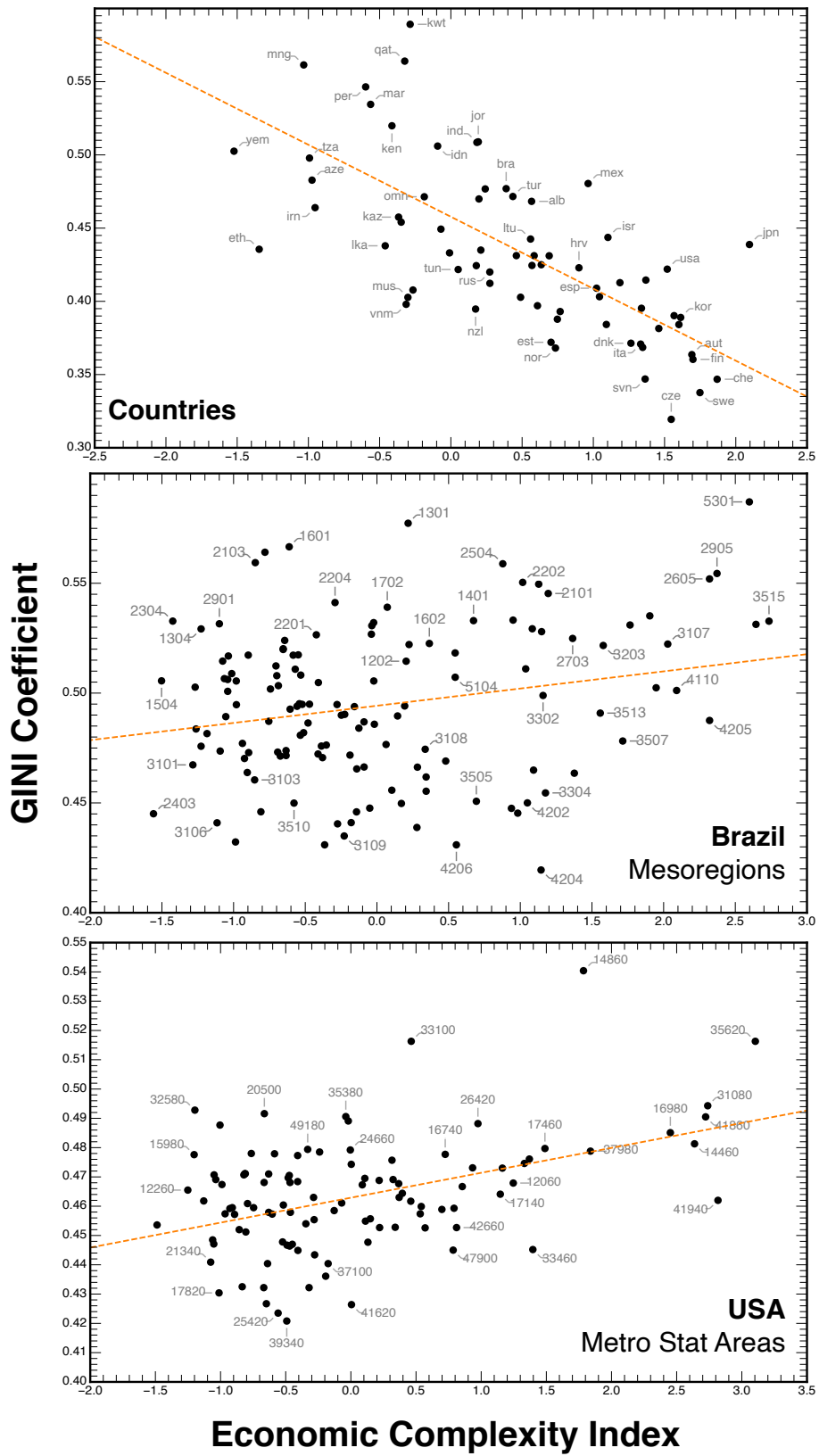

Figure 1. Simpson's paradox of the association between economic complexity and Gini inequality is negative on the national and positive on the regional level. Results report data for 2010.

# 2 Economic growth, productive specialization, and inequality

The debate on the relationship between economic development and inequality has a long tradition in economics. For instance, Simon Kuznet's (1955) argued that countries would naturally first see a rise and then a decline of within-country income inequality during the process of economic development, measured in terms of GDP. A pattern that would result in an inverted U-shaped relationship between GDP and inequality. The underlying reasoning is related to rural-urban migration patterns and institutional changes. During the first part of the development process, workers migrate to urban industrial sectors, leading to an increase in rural-urban inequality. Moreover, the profits of the company owners within the cities grow faster than the salaries of their workers. Thus, inequality rises. However, at some point, inequality falls again when a significant share of the workers moves to new industries with higher remuneration levels, but also drops due to the ensuing social tensions and institutional changes from electoral processes (such as democratization, redistribution policies, and welfare states).

However, the empirical evidence for the Kuznets curve is shaky at best. Recent data has shown that the Kuznets curve fails to hold on the country level when several Latin American countries are removed from the sample, and the upward side of the Kuznets curve has vanished in recent decades, as inequality in many low-income countries has increased (Stiglitz, 1996; Deininger & Squire, 1998; Palma, 2011). These findings undermine the empirical robustness of Kuznets' curve and indicate that GDP per capita is a measure of economic development that is insufficient to explain variations in income inequality. In fact, the type of products that countries or regions produce and their type of external embeddedness in the world trade network matter for their ability to promote inclusive growth and expected level of inequality. To depict this properly, new economic development measures, such as the economic complexity index (Hidaldo and Hausmann, 2009), are necessary. Moreover, it cannot simply be assumed that GDP growth necessarily automatically leads to a reduction of inequality, as high levels of inequality in several oil and other natural resources-rich countries illustrate. A more sophisticated perspective on the quality of economic growth, institutional changes, and path dependencies at the regional and national levels are necessary.

In that regard, development and institutional economists have shown that a focus on simple economic activities based on natural resources, cheap labor, and economies of scale has historically led to the emergence and replication of exploitative institutions over long periods (Engerman & Sokoloff, 1997). Moreover, several scholars have highlighted how the association between economic development and inequality might not only depend on the rate and stage of GDP growth, but also on the type of economic growth and sectoral structures, as well as the accompanying institutions and social policies (Stiglitz, 1996; Engerman & Sokoloff, 1997; Fields; 2002; Collier 2007; Hartmann et al., 2017).

Polarization and dependency theorists, such as Prebisch, Myrdal, Furtado, Frank, and Baran, criticized the Ricardian theory of comparative advantages lifting all boats, arguing that the free play of capitalist market forces would lead to core-periphery structures in the global economy and thus persistent structural inequality across countries (Meier and Seers, 1984; Bielschowsky, 2009). While the countries at the core diversify into high-value-added and manufactured products that provide their economy with high levels of productivity and comparatively good jobs for their citizens, countries in the periphery would specialize in simple products that are based on cheap labor and natural resources richness/exploitation. Moreover, within the countries, core-periphery structures also emerge across regions (Myrdal, 1957; Furtado, 1959; Diniz, 2009), and spatial agglomeration effects tend to perpetuate regional inequality. For instance, Myrdal (1957) argued that the free play of market forces might not lead to economic convergence between regions, as theorized by neoclassical approaches. Instead, cumulative causation leads to rich regions becoming even richer and poorer ones becoming poorer. This is due to backwash effects, such as the selective migration of young and educated, externalities of infrastructure for commerce, and capital movement towards economically more developed regions. These backward effects are, according to Myrdal (1957), stronger than potential spread effects, such as remittances and diffusion of technologies. Consequently, countries in the periphery may have high GDP industries in a few relatively rich and developed cities or resource-rich regions, but the remaining regions remain primarily poor. Moreover, the large surplus of cheap labor from poorer regions can also prevent higher (relative) salaries for workers in simple activities, such as urban services. Recent works on labor markets and technological changes also point to an increasing labor market polarization between high-tech activities and simple services in large cities (Alabdulkareem et al., 2018). Due to all these factors, the downwards slope of the Kuznets curve may not naturally occur, and it is even less likely to occur in the large urban centers of the developing economies.

# 3  New insights from economic complexity and relatedness research

Considering the above-mentioned insights from development economists, we can expect that more nuanced economic development measures, such as those focused on the variety and sophistication of products, would be able to provide deeper insights into the connection between economic development and income inequality. In other words, these measures can overcome the limitations of aggregate output measures, such as GDP. One of such measures is the Economic Complexity Index (ECI), introduced by Hidalgo and Hausmann (2009), which has been widely adopted in recent years to study the development of regions and countries (Hidalgo, 2021, Ferraz et al., 2021). ECI is a measure of the knowledge intensity of an economy that is expressed in the type of products the economy exports. Among the most complex export goods are sophisticated chemicals and machinery, whereas the least complex products are raw materials or simple agricultural products. Countries like Saudi Arabia, Chile, and Ghana rely heavily on a very limited number of simple and resource exploiting products, such as crude petroleum, copper, or cocoa beans, and therefore have a low ECI. Conversely, countries like Japan, South Korea, and Germany export a high diversity of very complex products, such as microchips, medicaments, and sophisticated car parts.

Since the seminal work of Hidalgo and Hausmann (2009), ECI — and related indicators, such as the Fitness Index, ECI+, or regional complexity measures — have been widely applied in the study of economic growth and diversification (Hausmann et al., 2014: Hidalgo, 2021). But only recently has empirical work explored the association between economic complexity indicators and inequality. Subsequent work has provided refined insights by exploring different measures of inequality and inclusive growth (such as human development, poverty, and jobs), different measures of economic complexity (such as ECI, ECI+, Fitness Index), considering intermediating factors, different countries, different temporal periods, and different levels of spatial aggregations (see next section). While research is still going on, a seemingly paradoxical result has emerged. While increasing ECI tends to be associated with lower income inequality at the country level, the relationship reverses at lower spatial aggregations, meaning it tends to be associated with an increased inequality at the regional level.

## 3.1 Economic complexity, inequality, and inclusive growth of nations

Using multivariate regression analysis, Hartmann et al. (2017) showed that ECI is a significant and negative predictor of income inequality at the country level. This relationship is robust when controlling for aggregate measures of income, institutions, and human capital. Scholars from different disciplines have previously argued that income inequality depends on various factors, from an economy's factor endowments, geography, and institutions to its historical trajectories, changes in technology, and returns to capital. Hartmann et al. (2017) argue that ECI represents a high-resolution expression of these factors, which co-evolve with the inclusiveness of its economy. Such co-evolution means that productive structures are not only associated with income and economic growth, but also with how income is distributed. For example, post-colonial economies specializing in agricultural or mineral products tend to have a more unequal distribution of political power, human capital, and wealth. Conversely, sophisticated products, like medical imaging devices or electronic components, are typically produced in complex economies with more inclusive institutions. Those economies employ more skilled workers, making firms more sensitive to their ability to attract and retain talent. Moreover, complex economies tend to be associated with a better distribution of political power (and lower levels of political capture of economic benefits and rent-seeking) than economies dependent on a limited number of resource-exploiting products.

Recent studies have further refined these findings by using different empirical methods, country sets, and time frames, as well as exploring different indicators for inclusive growth, such as human development, gender inequality, labor share, or the types of jobs in a country. Chu and Hoang (2020) use data for eighty-eight countries between 2002 to 2017 to show when economies reach/overcome a threshold level of education, government spending, and trade openness then higher ECI leads to a reduction of income inequality. However, higher ECI fails to reduce income inequality in countries with low education, ineffective government spending, and low economic openness. Lee and Wang (2021) also show that the association between ECI and inequality depends on other factors. They illustrate that country risk (i.e. economic risk, financial risk, and political risk) splits their panel data of 43 countries (from 1991 to 2016 into two groups. In countries with low risk, an increase in ECI is associated with more equal income distribution, while an improvement in ECI for in high-risk countries did not have a significant impact on their unequal income distribution. A study of Sepehrdoust et al. (2021) on middle-income countries also demonstrates that economic complexity reduces inequality above a certain threshold. Moreover, Lee and Vu (2019) use interaction terms to show that the ability of ECI to explain income inequality is mediated by the presence of high levels of education and good/inclusive institutions, suggesting that ECI only has an equalizing effect in the presence of good underlying labor market conditions. Lee and Vu (2019) show that human capital magnifies the negative correlation between complexity and inequality, arguing that countries endowed with better and improved human capital can enhance economic structures. This in turn reinforces the negative distributional effects of economic complexity, leaving them with a lower level of inequality. Additionally, Fawaz and Rahnama-Moghadamm (2019) also show that the type of trade partners impacts inequality and find that trade with more economically complex countries is correlated with reductions in income inequality.

Ben Saâd and Assoumou-Ella (2019) reveal that economic complexity has a positive effect on Gender Parity Index (GPI) in tertiary education (GPI), but not on primary and secondary education. Regarding labor market conditions, Barza et al. (2020) show that higher complexity industries and occupations in Brazil exhibit lower gender gaps in wages. Additionally, several contributions have pointed to a positive association between ECI and human development (Hartmann, 2014; Ferraz et al., 2018, Le Caous and Huarng, 2020), human capital in terms of secondary- and tertiary education (Zhu and Li, 2017), and health indicators (Vu, 2020) at the national level.

In respect to (un)employment, Gala et al. (2018) illustrate that in the long-run, economic complexity depends on the effort and the ability of countries to generate employment in manufacturing and sophisticated services sectors. Moreover, Adam et al. (2021) show that moving to higher levels of economic sophistication of exported goods leads to less unemployment and more employment, notably revealing that economic complexity does not induce job loss. Arif (2021) argues that high productive knowledge of workers increases their bargaining power, revealing a positive relationship between economic sophistication and labor share. However, this relationship is conditional on the level of human capital.

In sum, several works at the national level have shown evidence of a trend towards a negative relationship between economic complexity (ECI) and income inequality at higher levels of socioeconomic development, as well as a positive relationship between economic complexity and more knowledge-based jobs. However, they also suggest that these relationships depend on human capital and institutions that are present in the respective countries.

## 3.2 Economic complexity and inequality at the regional level

At the regional level, some studies have found a negative relationship between economic complexity (ECI) with income inequality or poverty at the regional level. For instance, Gao and Zhou (2018) report an negative effect of ECI on income inequality at the regional level in China. In addition, Zhu et al. (2020) show that ECI contributes to reducing income inequality in urban areas in China, but urban-rural inequality increases in regions with more complex export structures (i.e., greater ECI).

However, there are several good reasons why regional agglomeration effects can also lead to a positive association between economic complexity indicators and income inequality across and within regions of a country (Marco et al., 2022). For instance, the migration of low-skilled labor from poorer regions toward large urban centers can lead to a constant surplus of cheap labor and can result in high levels of income inequality in large cities, where both simple services and high-tech sectors co-locate and can create high levels of inequality. Moreover, population growth and housing prices can aggravate socioeconomic inequalities and ghettoization (Heinrich Mora et al., 2018; Youn et al., 2016). It is evident that there is an abysmal divide between wealthy business districts and large shantytowns in cities located in developing countries — such as Mumbai, Cape Town, Lima, or Sao Paulo — but also in cities in rich economies — such as London, Paris, New York, or San Francisco — where high inequality levels are also prevalent. Moreover, the rise of new tech and finance industries, labor market polarization, and new labor market monopsonies have further added to higher levels of inequality (Azar et al., 2022). In that regard, skill premiums and stock shares have made executives, lead developers, and other high-skilled occupations richer (Autor, 2014). In contrast, the labor share of industries (Autor et al., 2020) and the inflation-corrected wages have declined in many cities. Low to intermediate complex manufacturing activities may leave or avoid large cities due to negative externalities, traffic jams, and high prices; in consequence, reducing further job opportunities for the middle and lower-middle classes.

Additionally, it must be noted that regions of the same country do share similar institutions (such as public transfer, labor rights, languages, and culture) (Marco et al., 2022). Also, the workers from one region/city might not necessarily have a sufficient voice, votes, and power to change the national level institutions, laws, or taxes in order to adapt institutions according to their needs. Hence, a rise in ECI within a few large cities may not have the same power to promote more inclusive institutions than country-wise large-scale transformations in the productive structures and trade comparative advantages. Urban workers in simple services and manual activities can be relatively easy substituted by workers from other regions, thus reducing their power to demand higher wages, institutional changes, and distributive measures. At the same time, simple to intermediate manufacturing industries may have moved elsewhere, leaving a greater voice and power to high-skilled services, company owners, and richer strata of the large cities. This does not mean that cities cannot be places where ideas for social change are exchanged, social discontent erupts, and large-scale demonstrations can demand institutional changes. Quite the opposite, it is precisely this natural tendency towards inequality in large cities that often leads to social tension and triggers demands for institutional changes. But it also suggests that at the regional level, the migration and agglomeration effects may have an overall stronger effect on the co-evolution of high(er) level economic complexity and inequality, than the redistribution and equalizing effects of (formal) institutions and social policies on the country level.

Thus, due to agglomeration/polarization effects, the relationship between economic complexity (ECI) and inequality seemingly reverses at the regional level. Indeed, state-level data for Brazil shows a positive relationship between ECI and income inequality (and a small but negative effect at very high relative levels of ECI) (Morais et al., 2021). A similar pattern can be found in US counties (Sbardella et al., 2017) and in regions in Spain (Marco et al., 2022).

As seen in Figure 1, there is a (weak but) positive association between economic complexity and income inequality at the regional level in the US and Brazil. Additional research is necessary to include additional countries in this analysis. However, while the data shows a clear negative association between economic complexity and income inequality at the cross-sectional level of countries, the sign of the relationship is less clear and rather reverses at the regional level.

## 3.3 Ambiguous and dynamic effects of economic complexity

It must be noted that several papers have pointed to ambiguous effects of economic complexity on inequality. For instance, Lee and Vu (2019) use a cross-country sample to show a negative correlation between economic complexity and income inequality but also find evidence that complexity and inequality within countries increase together over time. The main reason for this behavior is that skilled workers find it easier to adapt to structural changes than low-skilled workers (Lee and Vu, 2019, Hartmann, 2014). Thus, in the initial phases of structural transformations, an increasing level of economic complexity (ECI) may also lead to increasing levels of inequality. Moreover, Lapatinas (2016) points to the ambiguous effects of structural change and economic diversification on human development, as outlined by Hartmann (2014). While economic diversification can increase people's choices and increase the demand for education, health, and other human development factors, it can also go along with economic polarization processes across regions in a country, temporary winners and losers, as well as increasing capability demands and problems of overchoice (Hartmann, 2014). This can balance out positive and negative effects (Lapatinas, 2016).

One reason for the ambiguous empirical results can stem from the stages in technology and industry life cycles, as well as where and when industries locate, grow, or decline in this process. In that context, the work of Carlota Perez (2003, 2007) on techno-economic paradigms implies that inequality within and across economies may first rise and then fall during the life cycle of new technological paradigms. New paradigms (such as the industrial revolution, the age of steam and railways, the age of steel and electricity, the emergence of mass production and automobiles, and the current information revolution/knowledge society) have led to 'opportunity explosions' focused on specific industries in specific countries and regions at early adoption stages. This, though, has recurrently led to financial bubbles and institutional crises. Only after installing more adequate institutions and infrastructure can the benefit of new technologies be more widely spread in different countries and parts of the economy. Now, the digital transformation seems to create superstar firms and a larger share of capital than labor (Autor et al., 2020), threaten to automatize simple jobs (Frey and Osborne, 2017), and lead to workplace and regional skills polarization (Alabdulkareem et al., 2018). But if history repeats, new infrastructure and more inclusive institutions may spread the benefits of technologies and new production abilities, and inequality may go down again.

Thereby, new waves of innovation can lead to structural transformations of the economies in terms of their productive specializations, institutions, and relative economic position, thus creating regional winners and losers (Schumpeter 1942; Aghion 2002; Mendez 2002; Perez, 2003; Boschma, 2021, Pinheiro et al., 2022). While some regions may win, others may decline. Nonetheless, there is a tendency for new activities to start first in larger cities that have the absorptive capacity, skilled labor, and innovative consumers present to enter new activities. These new growth dynamics in new (tech) sectors can raise inequality within cities and push less lucrative, unrelated industries out of cities. This though may not happen with complementary simple services that address the consumption needs of the workers of the new industries. This can result in rising inequality within and across regions/cities, at least in the early stages of new industry/technology life cycles. It is noteworthy that Kuznets' theory does not depict these constant changes in the technology and productive structures that can lead to more complicated dynamics between economic development and inequality over time. However, this trend toward new concentrations and inequality on the local level could indeed be partially overcome at some point by new laws and regulations, as well as efforts of social policies and redistribution on the national level, in addition to the diffusion of the new technologies and sectors towards other regions of the country.

## 3.4 Constraints and opportunities for inclusive growth

Economic complexity research has developed together with new methods to measure relatedness and, thus, revealed how related/similar economic activities are in terms of their productive capabilities, technologies, institutions, and other factors (Boschma, 2005; Hidalgo et al., 2018). Relatedness research helps to reveal the path-dependency of economic branching processes (Hidalgo et al. 2018; Hidalgo, 2021) and to identify constraints and opportunities of economies for smart and inclusive diversification (Hidalgo et al., 2007; Hidalgo et al., 2018; Balland et al., 2019). (Hartmann et al., 2016, 2017, 2019, 2021). Economies are seldom able to move into unrelated activities (Alshamsi et al., 2018; Hidalgo et al, 2018, Pinheiro et al., 2018, 2021), and developing related capabilities tends to be a more feasible and promising path to boost economic diversification and economic growth (Agosin, 2009; Hidalgo et al., 2018; Boschma and Iammarino, 2009; Boschma et al., 2012, 2014). However, the tendency of economies to move into related activities also implies the gravitation of less complex economies toward simple, cheap labor, and natural resources exploiting activities. In contrast, more complex economies tend to move and further specialize in more knowledge-based and inclusive activities (Hartmann et al., 2020, 2021, Pinheiro et al., 2021).

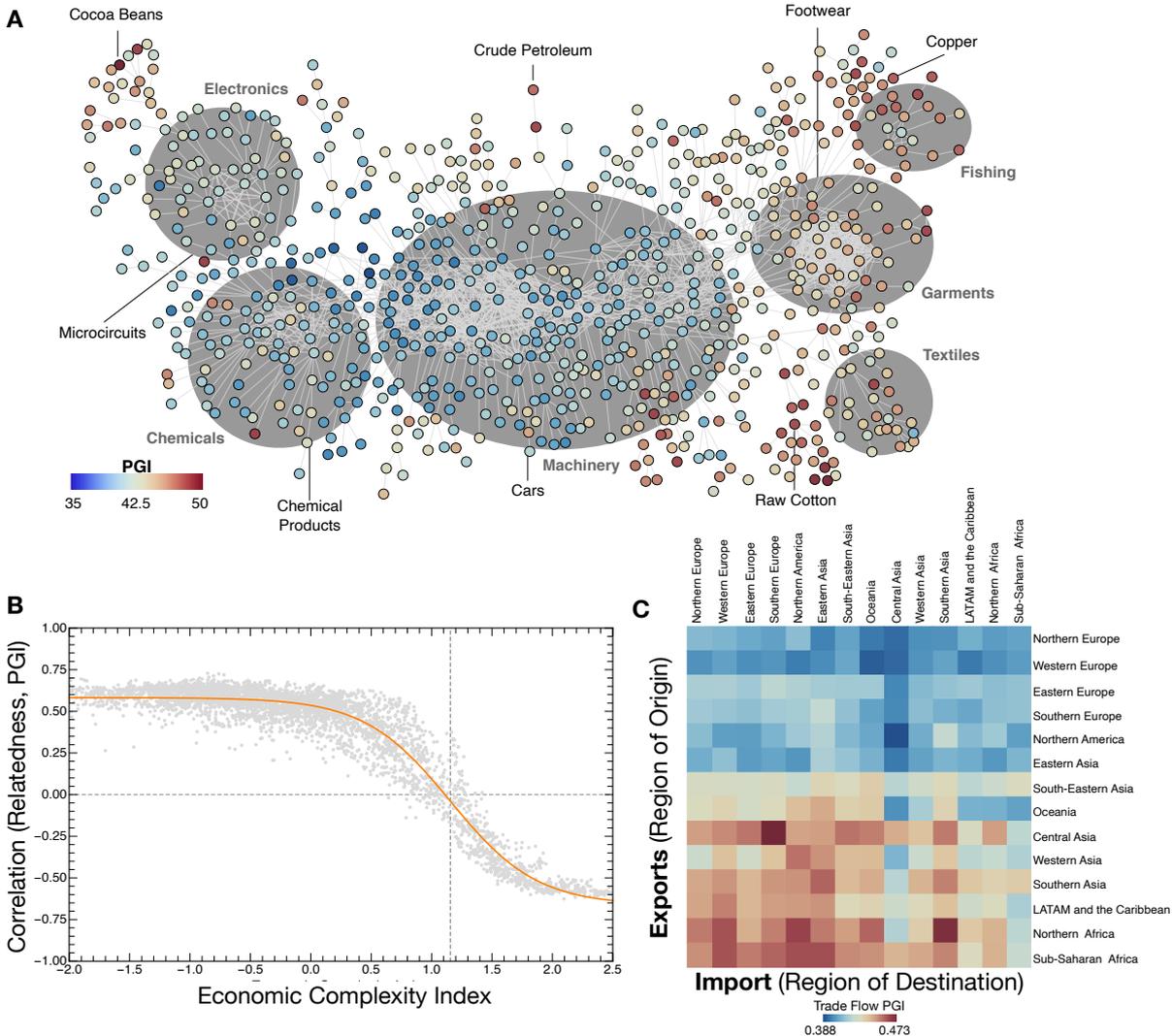

Figure 2. A) Product Space with products colored according to their level of PGI. Major Groups of products are highlighted. B) Proximity of countries to High PGI products as a function of their level of Economic Complexity Index, which we measure as the correlation between the relatedness and PGI of products for each country between 1980 and 2017. Each dot corresponds to a country in a year. C) Average PGI of major exports/imports between world regions, showing a core-periphery structure.

Overall, new methods in economic complexity research provide a suitable framework to confirm, adapt, and refine the qualitative ideas of polarization theories (e.g. development pioneers, structuralist economists, and world-system approaches) on structural imbalances across regions and countries, as well as the impact of productive specialization on the expected level of inequality and type of institutions in the production place. Recently these methods have been used to estimate the expected levels of income inequality associated with export goods (Hartmann et al., 2017), and thus for the locations in which these activities are present (Hartmann et al., 2016, 2019). The Product Gini Index (PGI) can be defined as the "average level of income inequality of a product's exporters, weighted by the importance of each product in a country's export basket" (Hartmann et al., 2017). Figure 2 shows the distribution of the PGIs in the product space. It reveals that products in the less connected periphery of the product space tend to be produced by countries with high levels of income inequality. Conversely, products in the highly connected center of the product space tend to be produced in countries with lower levels of inequality. PGI can be used to create a counterfactual level of income inequality for an economy, given its portfolio of activities (Hartmann et al., 2017: Hidalgo, 2021). Hartmann et al. (2016, 2020) have argued that the portfolio of productive activities is not the only factor explaining inequality (i.e. other factors such as tax systems, education, etc.); however, this portfolio significantly constrains the type of jobs available in an economy, distribution of income, and the possibilities for inclusive growth.

The large distance of developing countries from complex products creates severe development traps for them. A set of recent papers has illustrated that low complexity countries tend to be mainly close to simple products, while high complexity countries tend to be close to complex products (Pinheiro et al., 2018, Pinheiro et al., 2021, Hartmann et al., 2020; Hartmann et al., 2021). This large distance from complex products imposes a quiescence trap for developing economies that gravitate towards simple products (Hausmann and Hidalgo, 2010; Hartmann et al., 2021). For instance, Hartmann et al. (2019) show for the case of Paraguay that the most related product diversification opportunities would further reduce ECI of Paraguay and lead to further specialization in simple and high PGI products in the periphery of the product space. Paraguay would have chances to move into slightly more complex products gradually. Still, such steps would require effective coordination of policymakers, companies, scientists, and civil society. It is important to mention that many developing countries have managed to diversify into a set of simple products in the recent five to six decades. However, only very few economies have climbed the ladder of productive sophistication and transformed their economies towards a focus on products associated with low PGIs (Hartmann et al., 2019; 2021). These success cases combined smart industrial policies and accompanying social policies (Hartmann et al., 2021).

Research on structural transformation processes at different stages of productive sophistication also hints at why ECI tends to be associated with lower levels of income inequality at the national level. The most sophisticated countries tend to outsource activities with high PGI, such as those products based on cheap labor and somewhat exploitative institutions (Hartmann et al., 2020). This, for instance, includes outsourcing less value-added segments of textile industries, agricultural goods, or polluting industries. Instead, countries with high ECI tend to focus on sophisticated products associated with low PGIs, more inclusive institutions, and higher wages.

At the regional/city level, though, outsourcing these simple export industries does not have the same effect on reducing income inequality in large cities. This is because large cities tend to attract both low- and high-wage service industries. Low complexity export industries with low to intermediate salaries (e.g. textile industries or simple component assemblers) can face cost pressures (in large cities) and migrate to other regions. This can further increase income inequality within cities and lead to labor market polarization, while at the same time (slightly) raising the average complexity of the economic activities of cities. Thus, both the economic complexity index and income inequality may increase in cities.

## 4  Policy considerations

It is noteworthy that so far inequality measures have tended to focus on the pre-production factors (such as education) or the post-production factors (such as taxes and redistribution) (Rodrik and Stantcheva, 2021). Research on economic complexity and inequality shows us that the production stage is important as well. This is the case because it determines which and how many jobs are available in which industries. It makes a difference for the wage distribution, income inequality, and human development demands if an economy is mainly based on simple export goods or services (and some few executives and knowledge-based services industries) or if an economy is based on a significant number of sophisticated industries and high-skilled jobs.

Of course, even in highly complex economies, some simple services industries will still exist and can lead to substantial levels of inequality, especially in large cities and at the regional level. But overall, in countries with a high ECI, there is also a wider variety of job opportunities among more sophisticated industries and intermediate to high paid jobs available (Gala et al., 2018; Arif, 2021). Moreover, there is a higher likelihood for there to be more inclusive institutions as well as outsourcing much of the undesirable economic activities (with low wages and high pollution) to other countries. Having said that, more inclusive institutions are not a fully automatic process, and institutional differences can exist in economies with similar levels of economic complexity. While more complex industries tend to require more skilled labor and a learning society, differences in labor rights, taxes and redistribution, and education can exist. Thus, industrial policies and social-institutional policies may need to complement each other for the sake of inclusive growth (Stiglitz, 1996; Hartmann et al., 2016). Indeed, it has been shown that successful countries that have managed to catch up and leapfrog ahead economically, as seen with South Korea, Singapore, or Ireland, have combined smart industrial policies with complementary social policies, such as target education or inclusive housing programs (Hartmann et al., 2021).

Furthermore, emphasis on individual pre-production factors of inequality, such as education, can be important to promote social mobility, but the mere emphasis on education alone may not necessarily be enough to tackle the structural problem of inequality imposed by productive structures. For instance, a rise in average schooling years in poor agricultural regions does not automatically provide poor people with better job opportunities in their home regions. However, suppose they migrate to the slums in large cities of developing economies. This can even result in inequality rising, rather than falling. So, while the emphasis on more education can have multiple positive impacts, it arguably needs to be complemented by additional industrial policies that help to create better jobs.

Importantly new methods from economic complexity and relatedness research allow for a larger informational base on structural constraints for inequality reduction as well as the identification of inclusive growth opportunities (e.g., Hartmann et al., 2016, 2019; Bam and De Bruyne, 2019; Bam et al., 2021.). Thus, it also allows for more tailor-made industrial policies that consider the strengths and opportunities of each country, region, and industry to promote inclusive growth. In this context, so far research on industrial policies has tended to assume a positive trickle-down effect of manufacturing industries on poorer strata. However, little focus has been put on the more precise linkages between different industries and their impact on inequality and poverty reduction, particularly at the regional level in developing economies (Ferraz et al., 2021). In that regard, new methods from economic complexity can help to consider inequality, employment, and poverty within the industrial policy design.

Of course, research on the association between economic complexity and inequality is only at the beginning, and, for example, the dynamic co-evolution of complexity and inequality is not yet fully understood for different spatial levels, longer time periods, and types of economic activities. Additionally, it has not yet been fully understood when higher levels of inequality between and within cities are rather a sign of natural agglomeration effects in a dynamic economy (Pinheiro et al., 2022) and when the levels of inequality are so high that it hampers economic dynamics and inclusive growth. There are seemingly opposite forces at work that lead to the change in the association between development and inequality at different levels of spatial aggregation, from cities and regions to countries. The cyclical nature of technology and industry cycles of innovation can arguably imply that for countries to climb the ladder of development and execute all the necessary social and institutional transformations, a certain degree of regional inequality is a cost to be paid: a temporary trade-off of development that comes as a side effect of such transformations. It thus raises questions on whether inequality can be endemic or transformative: the former a result of unfit institutional settings, while the latter a result of the transformative forces (agglomeration, industrial and technological innovation cycles, etc.) that push countries towards development. In that sense, countries experience regional inequality for different reasons, and it is important to distinguish whether it rather hampers mid-to long-run economic diversification and sophistication, or if it is also a sign of structural transformation processes. We are only beginning to answer such questions empirically. Nonetheless, economic complexity research has allowed for significant progress in the empirical understanding of the effects of productive structure on the distribution of income and human development, as well as the identification of more or fewer desirable activities. This helps to build bridges between industrial and social policies for the sake of inequality reduction and inclusive growth. We now have the tools at hand to develop a more disaggregated and dynamic picture of the complex co-evolution of productive structures and inequality.

A rather linear conceptualization of development stages (including the Kuznets curve) may not be sufficient in depicting the constant changes implied by innovation to the productive structures and distribution of income and opportunities. It may also not be sufficient in revealing the apparently conflicting, though complementary, dynamical forces at the national and regional levels. Agglomeration effects and financial bubbles tend to push towards a negative association between economic complexity and inequality on the regional level. However, institutional changes toward more inclusive growth, a wider variety of job opportunities, and productive outsourcing can lead to a positive association at the national level. We argue that both phenomena co-evolve in dynamic economies.